          [2001/04/25 1.1 (PWD)]
\documentclass[letterpaper,twocolumn]{esapub} 

\usepackage{natbib,graphicx}

\def\la{\lower4pt\hbox{${\buildrel < \over \sim}$}}
\def\ga{\lower4pt\hbox{${\buildrel > \over \sim}$}}

\title{Coordinated Multiwavelength Observations and Spectral Variability
Modeling of Gamma-Ray Blazars}
\author{Markus B\"ottcher}
\affil{Department of Physics and Astronomy, Ohio University, 
Athens, OH 45701, USA}

\begin{document}


\maketitle

\begin{abstract}
Our recent progress on time-dependent modeling of the multiwavelength 
spectra and variability of blazars with leptonic and hadronic jet models 
is reviewed. Special emphasis is placed on X-ray spectral variability of 
low-frequency peaked (LBLs) and intermediate BL Lac objects (IBLs). 
As an example, recent observational and modeling results of an extensive
multiwavelength campaign of BL~Lacertae in 2000 are presented. It is
demonstrated how combined spectral and variability modeling of LBLs 
and IBLs can significantly constrain emission models and potential
variability scenarios. In the case of BL Lacertae, the variability
appears to be driven primarily by fluctuations in the spectral index
of the non-thermal, ultrarelativistic electron population in the jet.
Such constraints allow us to refine predictions of the intrinsic GeV 
$\gamma$-ray emission and the dominant electron cooling mechanism in 
these objects. 
\end{abstract}

\section{Introduction}

Blazars are the most extreme class of active galaxies known to date.
They have been observed in all wavelength bands --- from radio through
very-high energy (VHE) $\gamma$-ray frequencies. 66 blazars have been
identified as sources of $> 100$~MeV emission detected by the EGRET
telescope on board the {\it Compton Gamma-Ray Observatory}
\citep[CGRO;][]{hartman99}, and 6 blazars (Mrk~421: \citet{punch92,petry96}; 
Mrk~501: \citet{quinn96,bradbury97}; PKS 2155-314: \citet{chadwick99}; 
1ES~2344+514: \citet{catanese98}; 1H~1426+428: \citet{horan02,aharonian02}; 
1ES~1959+650: \citet{nishiyama99,holder03,aharonian03}) have now 
been detected at VHE $\gamma$-rays ($> 350$~GeV) by ground-based 
air \v Cerenkov telescopes. Many of the EGRET-detected $\gamma$-ray 
blazars appear to emit --- at least temporarily --- the bulk of their 
bolometric luminosity at $\gamma$-ray energies. Blazars exhibit 
variability at all wavelengths on time scales --- in some
cases --- down to less than an hour. VLBI radio observations and
monitoring often reveal one-sided kpc-scale jet structures, exhibiting 
apparent superluminal motion. The radio through optical emission 
from blazars often shows linear polarization, pointing towards a 
synchrotron origin.

\begin{figure*}
\includegraphics[width=\linewidth]{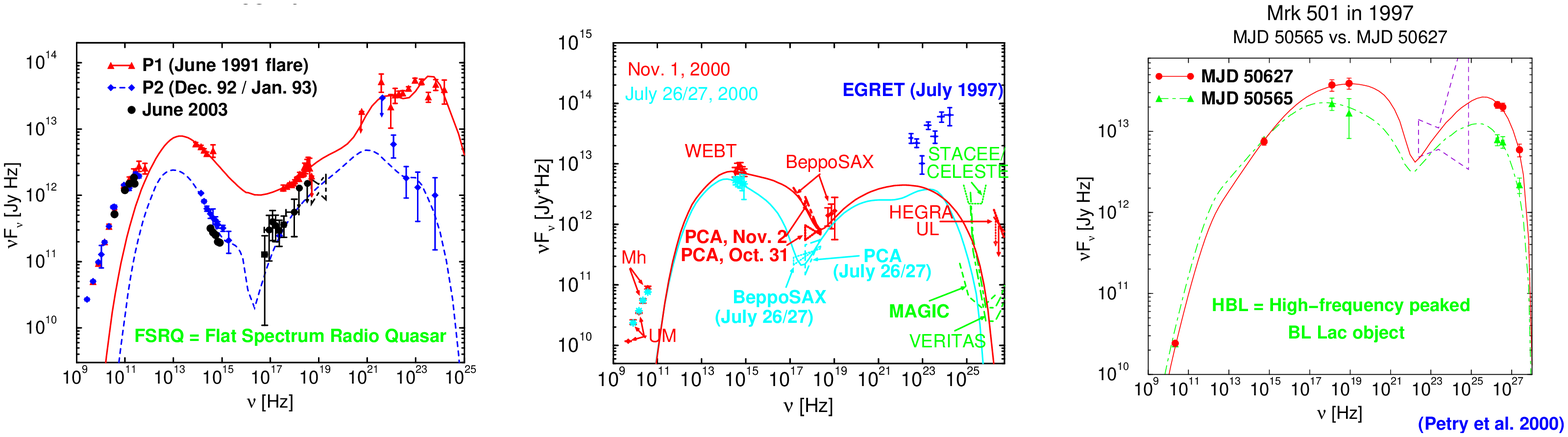}
\caption{Broadband spectra of the FSRQ 3C~279 
\citep[left panel;][]{hartman01,collmar04}, the LBL BL~Lacertae 
\citep[middle panel;][]{boettcher03}, and the HBL Mrk~501 
\citep{petry00}. 
For each object, two simultaneous SEDs at two different 
epochs are shown, illustrating the range of long-term 
spectral variability seen in these objects. For 3C~279, 
the results of recent multiwavelength observations in 2003,
including INTEGRAL \citep{collmar04} are also shown.}
\label{bbspectra}
\end{figure*}

\subsection{\label{obs_spectra}Blazar Spectra}

The broadband continuum spectra of blazars are dominated by
non-thermal emission and consist of at least two clearly 
distinct, broad spectral components. A sequence of sub-classes
of blazars can be defined through the peak frequencies and
relative $\nu F_{\nu}$ peak fluxes of those components, which 
also appear to be correlated with the overall bolometric 
luminosity of the sources \citep{fossati98}: In the case of 
flat-spectrum radio quasars (FSRQs), the low-frequency
(synchrotron) component extends from radio to optical/UV 
frequencies, with a peak frequency generally in the mm
or IR band; the high-frequency component extends from
X-rays through GeV $\gamma$-ray energies, with a 
$\nu F_{\nu}$ peak frequency corresponding to 
$\sim 10$~MeV --- 1~GeV. No FSRQ has so far been 
detected by ground-based air \v Cerenkov telescope
facilities at energies $> 100$~GeV, although in flaring states
the $\gamma$-ray $\nu F_{\nu}$ peak flux of FSRQs dominates 
over the low-frequency emission by up to $\sim 1$ order of 
magnitude. In the case of high-frequency peaked BL~Lac objects 
(HBLs), the low-frequency component often extends far into 
the X-rays, with peak frequencies ranging from the UV/soft 
X-ray to the hard X-ray regime \citep{pian98}, depending on 
the source and its state of activity; the high-energy component 
of HBLs extends from hard X-rays far into the VHE $\gamma$-ray 
regime. All blazars detected at VHE $\gamma$-ray energies to 
date are HBLs. In spite of extending to extremely high photon
energies, the $\nu F_{\nu}$ peak flux of the $\gamma$-ray component 
of HBLs is generally at most comparable to the spectral output in
the low-frequency component. In terms of their overall bolometric
luminosity, FSRQs appear to be several orders of magnitude more
powerful sources than HBLs.

Apparently intermediate between the FSRQs and the HBLs are
the low-frequency peaked BL~Lac objects (LBLs). The peak
of their low-frequency component is typically located at 
IR or optical wavelengths, their high-frequency component peaks 
typically at $\sim$~several GeV, and the $\gamma$-ray output
is of the order of or slightly higher than the level of the
low-frequency emission. Fig. \ref{bbspectra} shows a compilation
of broadband spectra of a typical FSRQ (3C279), an LBL (BL~Lacertae), 
and an HBL (Mrk~501), respectively. Each panel shows the respective 
object at two or three different observing epochs, in two different 
activity states.

\subsection{\label{variability}Blazar Variability}

\begin{figure*}
\includegraphics[width=\linewidth]{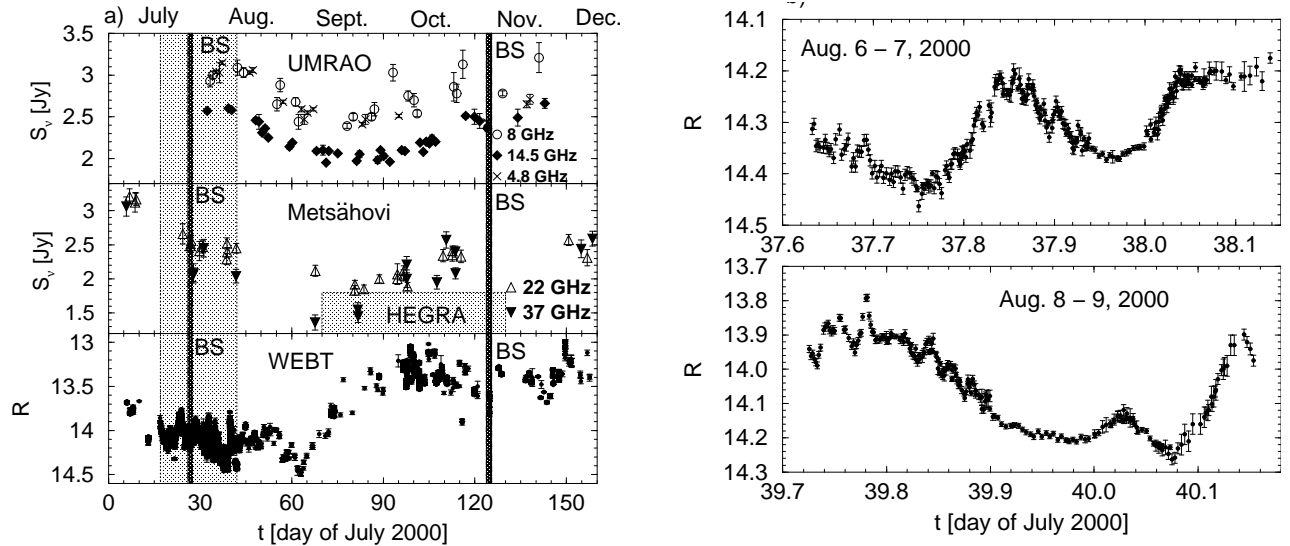}
\caption{Radio and optical light curves of the LBL BL~Lacertae 
during the multiwavelength campaign of 2000. The dark-shaded areas
in panel a) indicate the time segments of BeppoSAX observations.}
\label{bllac_lc}
\end{figure*}

Fig. \ref{bbspectra} already illustrates that in particular the
high-energy emission from blazars can easily vary by more than
an order of magnitude between different observing epochs. However,
variability has been observed on much shorter time scales, in some
extreme cases less than an hour \citep{gaidos96}. Fig. \ref{bllac_lc}
shows examples of light curves of the LBL BL~Lacertae, taken during
a broadband observing campaign carried out in the second half of
2000 \citep{boettcher03,villata02}. While the radio emission of blazars
is generally variable on time scales of weeks -- months (Fig. 
\ref{bllac_lc}a), the optical light curve shows significant 
variability on time scales down to $\sim 1.5$~hr (Fig.
\ref{bllac_lc}b). 

\begin{figure*}
\includegraphics[width=\linewidth]{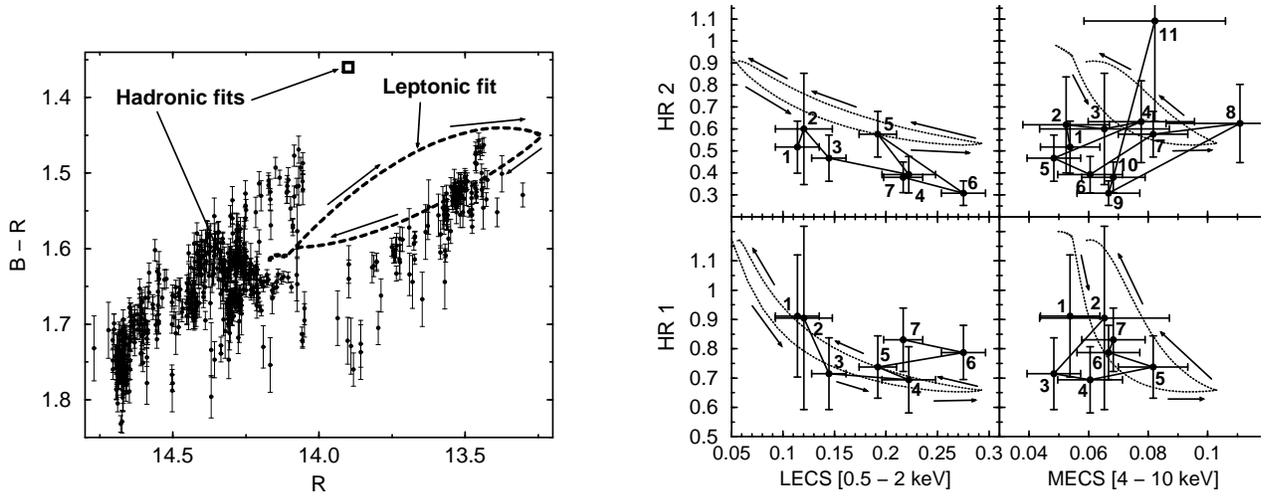}
\caption{Left panel (a): Optical color-magnitude relation observed
during the multiwavelength campaign on BL Lacertae in 2000
\citep{villata02,boettcher03,br04}.
Right panel (b): X-ray hardness-intensity diagram extracted from 
{\it Beppo}SAX observations. The curves indicate the spectral
variability patterns from the best-fit leptonic model of 
\citet{br04}.}
\label{spectralvar}
\end{figure*}

Often, both the optical and X-ray emission show characteristic
hardness-intensity correlations. Fig. \ref{spectralvar} illustrates
this for BL~Lacertae in 2000. Some HBLs (e.g., Mrk~421 and 
PKS~2155-304) have been observed to exhibit characteristic, clockwise 
loop structures \citep[``spectral hysteresis'';][]{takahashi96,kataoka00}, 
which can be interpreted as the synchrotron radiation signature of 
gradual injection and/or acceleration of ultrarelativistic 
electrons into the emitting region, and subsequent radiative 
cooling \citep{kirk98,gm98,kataoka00,kusunose00,lk00}.

In LBLs, the soft X-ray emission is also sometimes dominated by the
high-energy end of the synchrotron emission component, so similar
spectral hysteresis phenomena should in principle be observable. 
However, those objects are generally much fainter at X-ray energies
than their high-frequency peaked counterparts, making the extraction
of time-dependent spectral information an observationally very
challenging task. Fig. \ref{spectralvar} clearly illustrates that
the {\it Beppo}SAX observations of BL~Lacertae in 2000 revealed
evidence for spectral variability, but lacked the sensitivity
to clearly establish or rule out spectral hysteresis. Such a
measurement might require the new generation of X-ray
telescopes such as {\em Chandra} or {\em XMM-Newton}.

\section{Overview of Leptonic Jet Models of Blazars}

The high apparent bolometric luminosity combined with the short
variability time scales and the apparent superluminal motions of
individual jet components observed in many blazars, provide compelling
evidence that the nonthermal continuum is produced in emission 
regions of a typical size scale of $\sim$~a few light days or less, 
moving relativistically along a jet structure which is directed at a 
small angle with respect to our line of sight. The jets are most 
likely powered by accretion of circumnuclear matter onto a 
supermassive black hole of $10^8 \, M_{\odot} \la M_{\rm BH}
\la 10^{10} \, M_{\odot}$. The emission regions are characterized
by the presence of an ultrarelativistic population of nonthermal
electrons. Several scenarios have been proposed concerning the 
acceleration of such ultrarelativistic electrons, including 
impulsive injection near the base of the jet \citep{ds93,bms97}, 
individual shock waves propagating along the jet \cite{mg85}, 
relativistic particle acceleration at shear layers between 
a fast-moving inner jet and a slower moving outer jet 
\citep[e.g.,][]{so03} or internal shocks from the collisions 
of multiple shells of material ejected into the jet 
structure \citep{spada01}. Because of the difficulty of
constraining the acceleration mechanism and the composition
and spectral characteristics of the injected particle distribution
\citep[see, e.g.,][]{sm00,ob02,so03}, the time profile of injection
and injected particle spectra of ultrarelativistic electrons are
generally treated as free parameters in blazar modeling.

The nonthermal electrons are emitting synchrotron radiation, which 
is responsible for the low-frequency emission from radio to UV or 
even X-ray frequencies. Higher-frequency (X-ray and $\gamma$-ray)
emission is produced via Compton scattering processes. Possible 
target photon fields for Compton scattering are the synchrotron photons
produced within the jet \citep[the SSC process;][]{mg85,maraschi92,bm96}, 
or external photons (the EC process). Sources of external seed photons
include the UV -- soft X-ray emission from the disk --- either entering 
the jet directly \citep{dsm92,ds93} or after reprocessing in the broad 
line region (BLR) or other circumnuclear material \citep{sikora94,bl95,dss97}
---, jet synchrotron radiation reflected at the BLR \citep{gm96,bednarek98,bd98},
or the infrared emission from circumnuclear dust \citep{blaz00,arbeiter02}.
In the context of jet model invoking a significant deceleration of the
outflow along the jet, an important source of soft photons might 
also be provided by the synchrotron photons from slower portions
of the jet further downstream \citep{gk03}. 

In addition to these fundamental radiation processes, $\gamma\gamma$
absorption and pair production as well as synchrotron self absorption
have to be taken into account in order to build a self-consistent
blazar radiation model. Synchrotron self absorption is the reason
why some of the spectral fits shown in Fig. \ref{bbspectra} do not
reproduce the radio spectra of the respective sources: The emission
regions are optically thick at radio frequencies during the early
stages of propagation of the emission regions through the jet, where
the nonthermal electron population is sufficiently energetic to produce
ample high-energy radiation. 

As the emission region is moving relativistically outward, the
nonthermal electron population will evolve according to

$$
{\partial n_e (\gamma, t) \over \partial t} = - {\partial 
\over \partial \gamma} \left( \left[ {d\gamma \over dt} 
\right]_{\rm acc/loss, cont.} \, n_e [\gamma, t] \right) 
$$
\begin{equation}
+ q_e (\gamma, t) - p_e (\gamma, t) + Q_e (\gamma, t) - 
{n_{\rm e} (\epsilon, t) \over t_{\rm e, esc}},
\label{e_evolution}
\end{equation}
where $(d\gamma/dt)_{\rm acc/loss, cont.}$ denotes the
continuous energy losses due to radiative and adiabatic
cooling and energy gain due to acceleration processes,
$p_e$ and $q_e$ are the terms describing the population
and de-population of a given electron energy interval due 
to non-continuous energy loss processes (such as Compton
scattering, in particular in the Klein-Nishina regime), 
$Q_e$ describes the electron injection function, and 
$t_{\rm e, esc}$ is the escape time scale of nonthermal 
electrons. The evolution of the photon population in the 
emission region has to be solved simultaneously with the 
electron distribution, and is determined through

$$
{\partial n_{\rm ph} (\epsilon, t) \over \partial t} = 
\dot n_{\rm ph, em} (\epsilon, t) - \dot n_{\rm ph, abs}
(\epsilon, t) 
$$
\begin{equation}
+ q_{\rm ph} (\epsilon, t) - p_{\rm ph} 
(\epsilon, t) - {n_{\rm ph} (\epsilon, t) \over t_{\rm ph, esc}}
\label{ph_evolution}
\end{equation}
where now $\dot n_{\rm ph, em}$ and $\dot n_{\rm ph, abs}$
describe the fundamental emission and absorption processes,
$q_{\rm ph}$ and $p_{\rm ph}$ describe the scattering rates
into and out of a given photon energy bin, and $t_{\rm ph, esc}$
is the photon escape time scale. 

\subsection{\label{sp_results}Leptonic-Jet Spectral Modeling 
Results for Different Blazar Classes}

Various versions of the generic leptonic jet model described 
in the previous section have been used very successfully to 
model simultaneous broadband spectra of several FSRQs, LBLs, 
and HBLs. As more detailed spectral information has become 
available, the results of such broadband spectral modeling 
have now converged towards a rather consistent picture 
\citep{ghisellini98,kubo98}. The spectral sequence HBLs 
$\to$ LBLs $\to$ FSRQs appears to be related to an increasing 
contribution of the external Comptonization mechanism to the 
$\gamma$-ray spectrum. While most FSRQs are successfully 
modelled with external Comptonization models 
\citep[e.g.,][]{dss97,sambruna97,muk99,hartman01},
the broadband spectra of HBLs are consistent with pure SSC 
models \citep{mk97,pian98,petry00,kraw02}. BL~Lacertae, a LBL, 
appears to be intermediate between these two extremes, 
requiring an external Comptonization component to explain 
the EGRET spectrum \citep{madejski99,bb00}. One generally 
finds that HBLs require higher average electron 
energies and lower magnetic fields than LBLs and 
FSRQs. In most cases, the required Doppler boosting 
factors $D$ seem to be comparable for all types of objects,
although there have also been some results indicating an
extraordinarily high value of $D \ga 40$ for the HBL 
Mrk~501 \citep{kraw02}. Typical examples of broadband 
spectral fits consistent with this sequence of leptonic 
jet model parameters are shown in Fig. \ref{bbspectra}.
The occasional finding of very high Doppler factors, in
particular in some HBLs, has prompted \citet{gk03} to 
propose their model of a decelerating, stratified jet 
in which synchrotron photons from slower regions of 
the jet would serve as seed photons for Compton 
scattering, appearing slightly blue shifted in
the rest frame of the faster high-energy emission region
further upstream. Such a scenario could remove the need
for bulk Lorentz factors largely in excess of $\sim 10$
for HBLs.

\subsection{\label{var_modeling}Leptonic-Jet Modeling 
of Blazar Spectral Variability}

The generic blazar model described above is inherently time-dependent
and facilitates the modeling not only of the broadband SEDs, but also 
the detailed spectral variability of blazars. Studies of blazar spectral 
variability have been done in great detail for the case of pure SSC
models with electron cooling dominated by synchrotron losses, e.g., 
by \citet{kirk98,gm98,kataoka00,kusunose00,lk00}. In those papers, 
the spectral hysteresis observed in several HBLs was reproduced, 
significantly constraining model parameters beyond constraints
obtainable from pure spectral modeling. More recently, \citet{sikora01}
have extended these studies to an inhomogeneous jet model, also including
an external Compton component, so that the model would now also be applicable
to LBLs and quasars. They have applied their model to the special case of
3C~279. Based on their results, they interpreted the lack of a measurable 
time lag between the $\gamma$-ray and X-ray emission in that object 
\citep{hartman01b} as evidence that X-rays and $\gamma$-rays might be
produced co-spatially by electrons of similar energies. This, in turn,
provides evidence for two separate emission components being dominant
at X-rays and $\gamma$-rays. Most plausibly, this might indicate that
the X-ray emission is dominated by SSC emission, while the $\gamma$-rays
are dominated by external Compton emission.

\begin{figure}
\includegraphics[width=\linewidth]{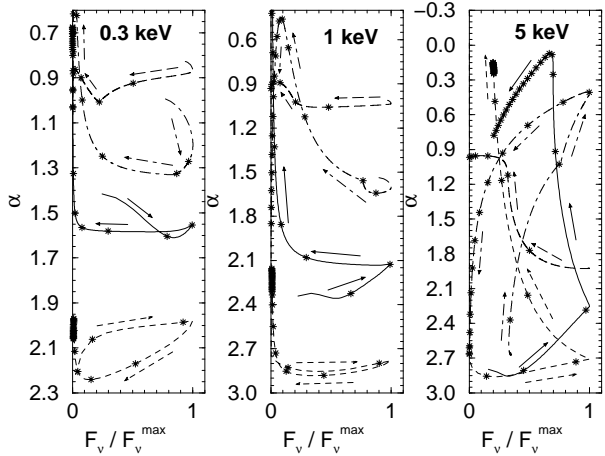}
\caption{Comparison of the tracks in the flux -- spectral-index
plane for different generic jet models, from synchrotron-dominated
(short-dashed) to SSC dominated (long-dashed). From \citet{bc02}.}
\label{bc02_hid}
\end{figure}

Time-dependent, homogeneous leptonic jet models in which radiation
mechanisms other than synchrotron may dominate have recently been
investigated in an analytical approach by \citet{cb02}, and with
detailed numerical simulations by \citet{bc02}. \citet{cb02} pointed 
out that a dominant contribution from SSC to the electron cooling
will produce a characteristic time-averaged synchrotron spectral
index of $\alpha = 3/2$ (energy spectral index), independent of the
injection index of relativistic electrons in the jet. In this case,
the cooling time scale of electrons radiating at a synchrotron
photon energy $E_{\rm sy}$ is expected to scale as $\tau_{\rm cool, SSC}
\propto E_{\rm sy}^{(q - 4)/2}$, where $q$ is the electron injection
spectral index \citep{boettcher03}. This differs characteristically 
from the synchrotron or EC dominated case in which $\tau_{\rm cool, sy} 
\propto E_{\rm sy}^{-1/2}$, independent of the electron injection
index $q$.

Detailed numerical simulations of the time-dependent emission
characteristics of homogeneous jet models with parameters specifically
chosen to be appropriate for low-frequency peaked and intermediate
BL~Lac objects have been done by \citet{bc02}. A key result of their
study was that a dominant SSC component would leave very obvious
imprints in the X-ray spectral variability of these objects, as
illustrated in Fig. \ref{bc02_hid}. In contrast, they found that
a moderate and even slightly dominant contribution from external
Compton scattering will have virtually no effect on the X-ray
spectral variability patterns. This might be a consequence of the
fact that the beaming pattern of external Commpton emission is more
strongly peaked in the forward direction than the synchrotron and
SSC components (which are assumed to be isotropic in the co-moving
frame of the emission region). Thus, even if the EC component is
dominating the $\gamma$-ray emission at MeV -- GeV energies, it may
only make a moderate contribution to the electron cooling rate.
The results of \citet{bc02} have been applied in detail to simultaneous
multiwavelength observations of W~Comae in 1998 \citep{bmr02} and 
BL~Lacertae in 2000 \citep{br04}.

\section{Overview of Hadronic Blazar Models}

While leptonic models deal with a relativistic e$^\pm$ plasma in the 
jet, in hadronic models the relativistic jet consists of a relativistic 
proton ($p$) and electron ($e^-$) component. In the following, a brief
summary of the Synchrotron-Proton Blazar (SPB-) model \citep{muecke03}
is given, as an example of a hadronic model that takes into account all 
the salient features of hadronic blazar jet models in general. 

Like in the leptonic model, the emission region in an AGN
jet moves 
relativistically along the jet axis which is closely aligned with 
our line of sight. Relativistic protons, whose particle density 
$n_p$ follows a power law spectrum $\propto \gamma_p^{-q_p}$ in 
the range $2\leq\gamma_p\leq\gamma_{\rm{p,max}}$, are injected 
instantaneously into a highly magnetized environment ($B =$~const. 
within the emission region), and are subject to energy losses due to 
proton-photon interactions (meson production and Bethe-Heitler pair 
production), synchrotron radiation and adiabatic expansion. The mesons 
produced in
photonmeson interactions always decay in astrophysical 
environments. However, they may suffer synchrotron losses before the 
decay, which is taken into account in this model.

\begin{figure}
\includegraphics[width=\linewidth]{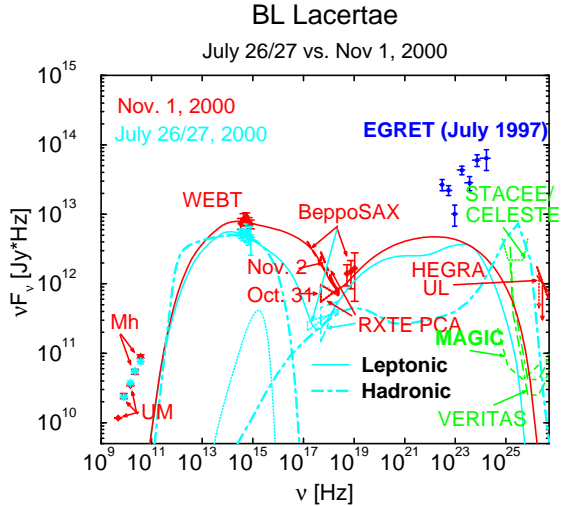}
\caption{Best-fit time-averaged leptonic (solid curves) and hadronic
(dot-dashed curves) models for BL~Lacertae in 2000 \citep{br04}.}
\label{bllacfit_lept_had}
\end{figure}

If the relativistic electrons are accelerated together with the 
protons at the same site, their injection spectrum shows most likely
the same spectral shape $\propto \gamma_e^{-q_e}$ with $q_e=q_p$.
The relativistic primary $e^{-}$ radiate synchrotron photons which
constitute the low-energy bump in the blazar SED, and serve as
the target radiation field for proton-photon interactions and
the pair-synchrotron cascade which subsequently develops. The 
SPB-model is designed for objects with a negligible external
target photon component, and hence suitable for BL~Lac objects.
The cascade redistributes the photon power to lower energies where
the photons eventually escape from the emission region. The
cascades can be initiated by photons from $\pi^0$-decay (``$\pi^0$ 
cascade''), electrons from the $\pi^\pm\to \mu^\pm\to e^\pm$ decay 
(``$\pi^\pm$ cascade''), $p$-synchrotron photons (``$p$-synchrotron 
cascade''), charged $\mu$-, $\pi$- and $K$-synchrotron photons 
(``$\mu^\pm$-synchrotron cascade'') and $e^\pm$ from proton-photon 
Bethe-Heitler pair production (``Bethe-Heitler cascade'').

Because ``$\pi^0$ cascades'' and ``$\pi^\pm$ cascades'' generate
rather featureless photon spectra \citep{muecke01,muecke03}, proton 
and muon synchrotron radiation and their reprocessed radiation turn 
out to be mainly responsible for the high energy photon output in 
blazars. The contribution from the Bethe-Heitler cascades is mostly
negligible. The low energy component is dominanted by synchrotron 
radiation from the primary $e^-$, with a small contribution of 
synchrotron radiation from secondary electrons (produced by the $p$- 
and 
$\mu^\pm$-synchrotron cascade). A detailed description of the 
model itself, and its implementation as a (time-independent) 
Monte-Carlo code, has been given in \cite{muecke01} and 
\cite{reimer04}.
This code has been used, e.g., to generate 
the spectral fits presented in \citet{br04} for BL~Lacertae 
in 2000 (see also Fig. \ref{bllacfit_lept_had}).

\section{Modeling Results for BL Lacertae in 2000}

Both the leptonic and hadronic models described in the previous
sections have been applied successfully to the data obtained
during the multiwavelength observing campaign on BL~Lacertae
in 2000 \citep{boettcher03}. The best time-averaged spectral 
fits are shown in Figs. \ref{bbspectra} and \ref{bllacfit_lept_had}.
The best-fit parameters for the leptonic and hadronic model fits
differ substantially in the following ways: (a) The overall jet
power required in the hadronic models is $\sim 2$ orders of magnitude
higher than for the leptonic models ($\sim 6 \times 10^{44}$~ergs~s$^{-1}$
vs. $\la 6 \times 10^{42}$~ergs~s$^{-1}$); (b) the magnetic fields
in hadronic jet models are a factor of $\sim 20$ higher than for
leptonic models ($B \sim 30$ -- 40~G vs. $\sim 2$~G), and the
bulk Lorentz factor of the emission region is a factor of $\sim 2$
lower for the hadronic models ($\sim 7$ -- 9 vs. $\sim 18$). 

Considering the time-averaged emission of BL~Lacertae in 2000, hadronic 
models predict a sustained level of multi-GeV -- TeV emission which 
should be detectable with second-generation atmospheric Cherenkov telescope 
systems like VERITAS, HESS, or MAGIC. In contrast, our leptonic model only 
predicts a peak flux exceeding the anticipated nominal MAGIC sensitivity 
during short flares; the accumulated fluence over observing time scales of 
several hours might not be sufficient for a significant detection. Thus,
a future VHE detection of BL~Lacertae would be a strong indication for 
hadronic processes being at work in this object.

\begin{figure}
\includegraphics[width=\linewidth]{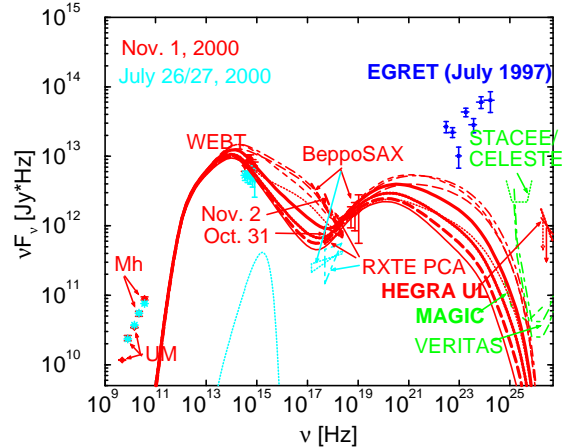}
\caption{Time-dependent model spectra from the best-fit leptonic
model for BL~Lacertae in 2000 \citep{br04}. The time sequence is:
thin solid $\to$ thin dotted $\to$ thin long-dashed $\to$ thin 
dot-dashed $\to$ thin short-dashed $\to$ thick solid $\to$ thick 
dotted $\to$ thick long-dashed, in equi-distant time steps of
$\Delta t_{\rm obs} \approx 1 hr$.}
\label{bllacfit_time}
\end{figure}

The detailed spectral evolution of our best-fit leptonic jet
simulation is shown in Fig. \ref{bllacfit_time}. In order to
determine the best-fit variability scenario, various generic
flaring scenarios had been investigated and compared to the
observed optical and X-ray spectral variability patterns (see
Fig. \ref{spectralvar}). Specifically, we had investigated
scenarios invoking a temporary increase in jet power, a flattening
of the electron injection spectral index, an increasing high-energy
cutoff $\gamma_2$ of the electron injection spectrum, and various
combination of these. Other parameter fluctuations could be ruled
out on the basis of general, analytical considerations, without 
detailed simulations. We found that the observed optical and 
X-ray spectral variability in BL~Lacertae in 2000 can be reproduced through
short-term fluctuations of only the electron injection spectral index, with 
all other parameters remaining unchanged. 

Remarkably, our time-dependent fits
indicated that an injection index 
larger than $q \sim 2.3$, even during the peak of an individual short-term 
flare, is required.
If the injection of ultrarelativistic electrons into 
the emitting volume
is caused by Fermi acceleration at relativistic shocks, 
detailed numerical
studies have shown that with fully developed turbulence 
in the downstream
region, a unique asymptotic index of $q \sim 2.2$ -- 2.3 
should be expected
\citep[e.g.,][]{achterberg01,gallant99}. However, recently 
\cite{ob02} have
shown that Fermi acceleration might lead to drastically 
steeper injection
spectra if the turbulence is not fully developed. 
Furthermore, depending on
the orientation of the magnetic field at 
the shock front, an abrupt steepening of the injection spectra may 
result if the shock transits from a subluminal to a superluminal 
configuration. In this context, our leptonic fit results may indicate 
that such predominantly geometric effects, may be the cause
of the 
rapid variability observed in BL~Lacertae. 

Our spectral-variability simulation predicted counter-clockwise spectral 
hysteresis at X-ray energies. Such hysteresis was not predicted in the 
specific SPB model fits presented in \citet{br04}, but could not clearly 
be ruled out either. Thus, sensitive spectral-hysteresis measurements of 
BL~Lacertae could possibly serve as a test of our modeling results and a 
secondary diagnostic to distinguish between leptonic and hadronic models, 
though, by itself, it would not be sufficient as a model discriminant. 

\citet{ravasio03} had previously noted the discrepancy between the 
time-averaged optical and X-ray spectra \citep{ravasio03} which could
not be joined smoothly by an absorbed power-law spectrum. They had
considered several possibilities to explain this discrepancy, including
additional particle populations, extreme Klein-Nishina effects on the 
electron cooling rates, and/or anomalies in the intergalactic absorption.
Our successful modeling of the observed time-dependent flux and hardness 
values at optical and X-ray frequencies in the framework of a leptonic model
effectively removes the need for such additional assumptions and indicates
that the time-averaging involved in compiling the detailed broadband
spectral energy distribution may be the cause of this apparent discrepancy.

\section*{Acknowledgments}

This work was partially supported by NASA through INTEGRAL GO
Program Grants no. NAG~5-13205 and NAG~5-13684.


\begin{thebibliography}{}

\bibitem[Achterberg et al., 2001]{achterberg01}Achterberg, A., Gallant, Y. A.,
Kirk, J. G., \& Guthmann, A. W., 2001, MNRAS, 328, 393

\bibitem[Aharonian et al.(2002)]{aharonian02}
Aharonian, F. A., et al., 2002, A\&A, 384, L23

\bibitem[Aharonian et al.(2003)]{aharonian03}
Aharonian, F. A., et al., 2003, A\&A, 406, L9

\bibitem[Arbeiter et al.(2002)]{arbeiter02}
Arbeiter, C., Pohl, M., \& Schlickeiser, R., 2002, A\&A, 
386, 415

\bibitem[Bednarek(1998)]{bednarek98}
Bednarek, W., A\&A, 342, 69

\bibitem[Blandford \& Levinson(1995)]{bl95}
Blandford, R. D., \& Levinson, A., 1995, ApJ, 441, 79

\bibitem[Bla$\dot{\rm z}$ejowski et al.(2000)]{blaz00}
Bla$\dot{\rm z}$ejowski, M., et al., 2000, ApJ, 545, 107

\bibitem[Bloom \& Marscher(1996)]{bm96}
Bloom, S. D., \& Marscher, A. P., 1996, ApJ, 461, 657

\bibitem[B\"ottcher et al.(1997)]{bms97}
B\"ottcher, M., Mause, h., \& Schlickeiser, R., 1997, A\&A,
324, 395

\bibitem[B\"ottcher \& Bloom(2000)]{bb00}
B\"ottcher, M., \& Bloom, S. D., 2000, AJ, 119, 469

\bibitem[B\"ottcher \& Chiang(2002)]{bc02}
B\"ottcher, M., \& Chiang, J., 2002, ApJ, 581, 127

\bibitem[B\"ottcher \& Dermer(1998)]{bd98}
B\"ottcher, M., \& Dermer, C. D., 1998, ApJ, 501, L51

\bibitem[B\"ottcher et al.(2003)]{boettcher03}
B\"ottcher, M., et al., 2003, ApJ, 598, 847

\bibitem[B\"ottcher et al.(2002)]{bmr02}
B\"ottcher, M., Mukherjee, R., \& Reimer, A., 2002,
ApJ, 581, 143

\bibitem[B\"ottcher \& Reimer(2004)]{br04}
B\"ottcher, M., \& Reimer, A., 2004, ApJ, submitted

\bibitem[Bradbury et al.(1997)]{bradbury97}
Bradbury, S. M., 1997, A\&A, 320, L5

\bibitem[Catanese et al.(1998)]{catanese98}
Catanese, M., et al., 1998, ApJ, 501, 616

\bibitem[Chadwick et al.(1999)]{chadwick99}
Chadwick, P. M., 1999, ApJ, 513, 161

\bibitem[Chiang \& B\"ottcher(2002)]{cb02}
Chiang, J., \& B\"ottcher, M., 2002, ApJ, 564, 92

\bibitem[Collmar et al.(2004)]{collmar04}
Collmar, W., et al., 2004, these proceedings

\bibitem[Dermer et al.(1992)]{dsm92}
Dermer, C. D., Schlickeiser, R., \& Mastichiadis, A.,1992,
A\&A, 256, L27

\bibitem[Dermer \& Schlickeiser(1993)]{ds93}
Dermer, C. D., \& Schlickeiser, R., 1993, ApJ, 416, 458

\bibitem[Dermer et al.(1997)]{dss97}
Dermer, C. D., Sturner, S. J., \& Schlickeiser, R., 1997,
ApJS, 109, 103

\bibitem[Fossati et al.(1998)]{fossati98}
Fossati, G., 1998, MNRAS, 299, 433

\bibitem[Gaidos et al.(1996)]{gaidos96}
Gaidos, J. A., et al., 1996, Nature, 383, 319

\bibitem[Gallant et al., 1999]{gallant99}Gallant, Y. A., Achterberg, A., 
\& Kirk, J. G., 1999, A\&AS, 138, 549

\bibitem[Georganopoulos \& Kazanas(2003)]{gk03}
Georganopoulos, M., \& Kazanas, D., 2003, ApJ, 594, L27

\bibitem[Georganopoulos \& Marscher(1998)]{gm98}
Georgenopoulos, M., \& Marscher, A. P., 1998, ApJ, 506, L11

\bibitem[Ghisellini \& Madau(1996)]{gm96}
Ghisellini, G., \& Madau, P., 1996, MNRAS, 280, 67

\bibitem[Ghisellini et al.(1998)]{ghisellini98}
Ghisellini, G., et al., 1998, MNRAS, 301, 451

\bibitem[Hartman et al.(1999)]{hartman99}
Hartman, R. C., et al., 1999, ApJS, 123, 79

\bibitem[Hartman et al.(2001a)]{hartman01}
Hartman, R. C., et al., 2001a, ApJ, 553, 683

\bibitem[Hartman et al.(2001b)]{hartman01b}
Hartman, R. C., et al., 2001b, ApJ, 558, 583

\bibitem[Holder et al.(2003)]{holder03}
Holder, J., et al., 2003, ApJ, 583, L9

\bibitem[Horan et al.(2002)]{horan02}
Horan, D., et al., 2002, ApJ, 571, 753

\bibitem[Kataoka et al.(2000)]{kataoka00}
Kataoka, J., et al., 2000, ApJ, 528, 243

\bibitem[Kirk et al.(1998)]{kirk98}
Kirk, J. G., Rieger, F. M., \& Mastichiadis, A., A\&A, 333, 452

\bibitem[Krawczynski et al.(2002)]{kraw02}
Krawczynski, H., Coppi, P. S., \& Aharonian, F. A., 2002, MNRAS,
336, 721

\bibitem[Kubo et al.(1998)]{kubo98}
Kubo, H., et al., 1998, ApJ, 504, 693

\bibitem[Kusunose et al.(2000)]{kusunose00}
Kusunose, M., Takahara, F., \& Li, H., 2000, ApJ, 536, 299

\bibitem[Li \& Kusunose(2000)]{lk00}
Li, J., \& Kusunose, M., 2000, ApJ, 536, 729

\bibitem[Madejski et al.(1999)]{madejski99}
Madejski, G., et al., 1999, ApJ, 521, 145

\bibitem[Maraschi et al.(1992)]{maraschi92}
Maraschi, L., Ghisellini, G., \& Celotti, A., 1992, 397, L5

\bibitem[Marscher \& Gear(1985)]{mg85}
Marscher, A. P., \& Gear, W. K., 1985, ApJ, 298, 114

\bibitem[Mastichiadis \& Kirk(1997)]{mk97}
Mastichiadis, A., \& Kirk, J. G., 1997, A\&A, 320, 19

\bibitem[M\"ucke \& Protheroe, 2001]{muecke01}M\"ucke, A., \& Protheroe, R. J.,
2001, Astropart. Phys., 15, 121

\bibitem[M\"ucke et al., 2003]{muecke03}M\"ucke, A., Protheroe, R. J.,
Engel, R., Rachen, J. P., \& Stanev, T., 2003, Astropart. Phys., 18, 593

\bibitem[Mukherjee et al.(1999)]{muk99}
Mukherjee, R., et al., 1999, ApJ, 527, 132

\bibitem[Nishiyama et al.(1999)]{nishiyama99}
Nishiyama, T., et al., in Proc. of the 26th ICRC, 3, 370

\bibitem[Ostrowski \& Bednarz(2002)]{ob02}
Ostrowski, M., \& Bednarz, J., 2002, A\&A, 394, 1141

\bibitem[Petry et al.(1996)]{petry96}
Petry, D., et al., 1996, A\&A, 311, L13

\bibitem[Petry et al.(2000)]{petry00}
Petry, D., et al., 2000, ApJ, 536, 742

\bibitem[Pian et al.(1998)]{pian98}
Pian, E., 1998, ApJ, 492, L17

\bibitem[Punch et al.(1992)]{punch92}
Punch, M., et al., 1992, Nature, 358, 477

\bibitem[Quinn et al.(1996)]{quinn96}
Quinn, J., et al., ApJ, 456, L83

\bibitem[Ravasio et al., 2003]{ravasio03}Ravasio, M., Tagliaferri, G.,
Ghisellini, G., Tavecchio, F., B\"ottcher, M., \& Sikora, M., 2003, A\&A, 
408, 479

\bibitem[Reimer et al., 2004]{reimer04}Reimer, A., Protheroe, R. J.,
\& Donea, A.-C., 2004, A\&A, in press

\bibitem[Sambruna et al.(1997)]{sambruna97}
Sambruna, R., et al., 1997, ApJ, 474, 639

\bibitem[Sikora \& Madejski(2000)]{sm00}
Sikora, M., \& Madejski, G., 2000, ApJ, 534, 109

\bibitem[Sikora et al.(2001)]{sikora01}
Sikora, M., et al., 2001, ApJ, 554, 1;
Erratum: ApJ, 561, 1154 (2001)

\bibitem[Spada et al.(2001)]{spada01}
Spada, M., MNRAS, 325, 1559

\bibitem[Sikora et al.(1994)]{sikora94}
Sikora, M., Begelman, M. C., \& Rees, M. J., 1994, ApJ, 421, 153

\bibitem[Stawarz \& Ostrowski(2003)]{so03}
Stawarz, L., \& Ostrowski, M., 2003, New Astron. Rev., 
47, 6-7, 521

\bibitem[Takahashi et al.(1996)]{takahashi96}
Takahashi, T., et al., 1996, ApJ, 470, L89

\bibitem[Villata et al.(2002)]{villata02}
Villata, M., et al., 2002, A\&A, 390, 407

\end{thebibliography}

\end{document}